\documentstyle[12pt,worldsci,epsf]{article}
\begin{document}

%%%%%%%%%%%%%%%%%%%%%%%%

%\preprint{\font\fortssbx=cmssbx10 scaled \magstep2
%\hfill$\vcenter \hbox{\bf IFUSP-P 1170}
%                \hbox{\bf hep-ph/9508342}
%}$}

{\font\fortssbx=cmssbx10 scaled \magstep2
\hfill \hbox{\bf IFUSP-P 1170\ and\ hep-ph/9508342} }

\vspace{.2in}

\title{Bounds on Scalar Leptoquarks from the LEP Data}

\author{Oscar J.\ P.\ \'Eboli \thanks{ E-mail:
    eboli@snfma1.if.usp.br (internet) or 47602::EBOLI (decnet).} \\ 
{\em Instituto de F\'{\i}sica, Universidade de S\~ao Paulo, \\ Caixa
  Postal 20516, 01452-990 S\~ao Paulo, Brazil.} }

\maketitle

\hfuzz=25pt

\begin{abstract}
  We obtain the constraints on scalar leptoquarks coming from
  radiative corrections to $Z$ physics.  We perform a global fitting
  to the LEP data including the contributions of the most general
  effective Lagrangian for scalar leptoquarks, which exhibits the
  $SU(2)_L \times U(1)_Y$ gauge invariance. Our bounds on leptoquarks
  that couple to the top quark are much stronger than the ones
  obtained from low energy experiments.
\end{abstract}

%******************************************************************************
\section{Introduction}

A large number of extensions of the SM predict the existence of color
triplet particles carrying simultaneously leptonic and baryonic
number, the so-called leptoquarks. Leptoquarks are present in models
that treat quarks and leptons on the same footing, such as composite
models \cite{comp}, grand unified theories \cite{gut}, technicolor
models \cite{tech}, and superstring-inspired models \cite{rizzo}.

Since leptoquarks are an undeniable signal for physics beyond the SM,
there have been several direct searches for them in accelerators.  At
the CERN Large Electron-Positron Collider (LEP), the experiments
established a lower bound $M_{LQ} \gtrsim 45$--$73$ GeV for scalar
leptoquarks \cite{lep}. On the other hand, the search for scalar
leptoquarks decaying into an electron-jet pair in $p\bar{p}$ colliders
constrained their masses to be $M_{LQ} \gtrsim 113$ GeV \cite{ppbar}.
Furthermore, the experiments at the DESY $ep$ collider HERA
\cite{hera} place limits on their masses and couplings, leading to
$M_{LQ} \gtrsim 92-184$ GeV depending on the leptoquark type and
couplings. There have also been many studies of the possibility of
observing leptoquarks in the future $pp$ \cite{fut:pp}, $ep$
\cite{buch,fut:ep}, $e^+e^-$ \cite{fut:ee}, $e\gamma$ \cite{fut:eg},
and $\gamma\gamma$ \cite{fut:gg} colliders.

In this work we study the constraints on scalar leptoquarks that can
be obtained from their contributions to the radiative corrections to
the $Z$ physics. We evaluated the one-loop contribution due to
leptoquarks to all LEP observables and made a global fit in order to
extract the 95\% confidence level limits on the leptoquarks masses and
couplings \cite{nois}. The most stringent limits are for leptoquarks
that couple to the top quark. Therefore, our results turn out to be
complementary to the low energy bounds \cite{leurer,davi} since these
constrain more strongly first and second generation leptoquarks.

The masses and couplings of leptoquarks are constrained by low-energy
experiments, since the leptoquarks induce two-lepton--two-quark
effective interactions, for energies much smaller than their masses
\cite{leurer,davi}. The processes that lead to strong limits are:

$\bullet$ Leptoquarks can give rise to flavor changing neutral current
(FCNC) processes if they couple to more than one family of quarks or
leptons \cite{shanker,fcnc}. In order to avoid strong bounds from
FCNC, we assumed that the leptoquarks couple to a single generation
of quarks and a single one of leptons. However, due to mixing effects
on the quark sector, there is still some amount of FCNC \cite{leurer}
and, therefore, leptoquarks that couple to the first two generations
of quarks must comply with some low-energy bounds \cite{leurer}.

$\bullet$ The analyses of the decays of pseudoscalar mesons, like the
pions, put stringent bounds on leptoquarks unless their coupling is
chiral -- that is, it is either left-handed or right-handed
\cite{shanker}.

$\bullet$ Leptoquarks that couple to the first family of quarks and
leptons are strongly constrained by atomic parity violation
\cite{apv}.  In this case, there is no choice of couplings that avoids
the strong limits.

It is interesting to keep in mind that the low-energy data constrain
the masses of the first generation leptoquarks to be bigger than
$0.5$--$1$ TeV when the coupling constants are equal to the
electromagnetic coupling $e$ \cite{leurer}.

The bounds on scalars leptoquarks coming from low-energy and $Z$
physics exclude large regions of the parameter space where the new
collider experiments could search for these particles, however, not
all of it \cite{fut:pp,fut:ep,fut:ee,fut:eg,fut:gg}.  Notwithstanding,
we should keep in mind that nothing substitutes the direct
observation.

%******************************************************************************

\section{Effective Interactions and Analytical Expressions}
\label{l:eff}

A natural hypothesis for theories beyond the SM is that they exhibit
the gauge symmetry $SU(2)_L \times U(1)_Y$ above the symmetry breaking
scale $v$. Therefore, we imposed this symmetry on the leptoquark
interactions.  In order to avoid strong bounds coming from the proton
lifetime experiments, we required baryon ($B$) and lepton ($L$) number
conservation.  The most general effective Lagrangian for leptoquarks
satisfying the above requirements and electric charge and color
conservation is \cite{buch}
\begin{eqnarray}
{\cal L}_{{eff}} &  & = {\cal L}_{F=2} ~+~ {\cal L}_{F=0} 
\; , 
\nonumber \\
{\cal L}_{F=2}  &  & = \left ( g_{{1L}}~ \bar{q}^c_L~ i \tau_2~ \ell_L + 
g_{{1R}}~ \bar{u}^c_R~ e_R \right )~ S_1 
+ \tilde{g}_{{1R}}~ \bar{d}^c_R ~ e_R ~ \tilde{S}_1
+ g_{3L}~ \bar{q}^c_L~ i \tau_2~\vec{\tau}~ \ell_L \cdot \vec{S}_3 
~ ,
\label{lag:fer}
\label{eff} \\
{\cal L}_{F=0}  &  & = h_{{2L}}~ R_2^T~ \bar{u}_R~ i \tau_2 ~ \ell_L 
+ h_{{2R}}~ \bar{q}_L  ~ e_R ~  R_2 
+ \tilde{h}_{{2L}}~ \tilde{R}^T_2~ \bar{d}_R~ i \tau_2~ \ell_L
\; ,
\nonumber
\end{eqnarray}
where $F=3B+L$, $q$ ($\ell$) stands for the left-handed quark (lepton)
doublet, and $u_R$, $d_R$, and $e_R$ are the singlet components of the
fermions. We denote the charge conjugated fermion fields by
$\psi^c=C\bar\psi^T$ and we omitted in (\ref{lag:fer}) the flavor
indices of the couplings to fermions and leptoquarks. The leptoquarks
$S_1$ and $\tilde{S}_1$ are singlets under $SU(2)_L$ while $R_2$ and
$\tilde{R}_2$ are doublets, and $S_3$ is a triplet.  Furthermore, we
assumed in this work that the leptoquarks belonging to a given
$SU(2)_L$ multiplet are degenerate in mass, with their mass denoted by
$M$.

Local invariance under $SU(2)_L \times U(1)_Y$ implies that
leptoquarks also couple to the electroweak gauge bosons. To obtain the
couplings to $W^\pm$, $Z$, and $\gamma$, we substituted $\partial_\mu$
by the electroweak covariant derivative ($D_\mu$) in the leptoquark
kinetic Lagrangian:
\begin{equation}
D_\mu \Phi = \left [ \partial_\mu -
i \frac{e}{\sqrt{2} s_W} \left ( W_\mu^+ T^+ + W_\mu^- T^- \right ) - 
i e Q_Z Z_\mu
+ i e Q^\gamma A_\mu \right ] \Phi \; ,
\end{equation}
where $\Phi$ stands for the leptoquarks fields, $Q^\gamma$ is the
electric charge matrix of the leptoquarks, $s_W$ is the sine of the
weak mixing angle, and the $T$'s are the generators of $SU(2)_L$ for
the representation of the leptoquarks. The weak neutral charge is $Q_Z
= (T_3 - s_W^2 Q^\gamma)/s_W c_W$.

We employed the on-shell-renormalization scheme, adopting the
conventions of Ref.\ [20]. We used as inputs the fermion
masses, $G_F$, $\alpha_{{em}}$, and the $Z$ mass, and the
electroweak mixing angle being a derived quantity that is defined
through $\sin^2 \theta_W = s_W^2 \equiv 1 - M^2_W / M^2_Z$. We
evaluated the loops integrals using dimensional regularization and we
adopted the Feynman gauge to perform the calculations.

Close to the $Z$ resonance, the physics can be summarized by the
effective neutral current
\begin{equation}
J_\mu =  \left ( \sqrt{2} G_\mu M_Z^2 \rho_f 
\right )^{1/2} \left [ \left ( I_3^f - 2 Q^f s_W^2 \kappa_f \right )
\gamma_\mu - I_3^f \gamma_\mu \gamma_5 \right ] \; ,
\label{form:nc}
\end{equation}
where $Q^f$ ($I_3^f$) is the fermion electric charge (third component
of weak isospin).  The form factors $\rho_f$ and $\kappa_f$ have universal
contributions, {\em i.e.} independent of the fermion species, as well
as non-universal parts:
\begin{eqnarray}
 \rho_f   & = &  1 + \Delta \rho_{{univ}} + 
\Delta \rho_{{non}} \; , \\
\kappa_f  & = &  1 + \Delta \kappa_{{univ}} + 
\Delta \kappa_{{non}} \; .
\end{eqnarray}

Leptoquarks can affect the physics at the $Z$ pole through their
contributions to both universal and non-universal corrections. The
universal contributions can be expressed in terms of the
unrenormalized vector boson self-energy ($\Sigma$) as
\begin{eqnarray}
\Delta \rho^{LQ}_{{univ}}(s)  &=&  
-\frac{\Sigma^Z_{LQ}(s)-\Sigma^Z_{LQ}(M_Z^2)}{s-M_Z^2} 
+\frac{\Sigma^Z_{LQ}(M_Z^2)}{M_Z^2}
-\frac{\Sigma^W_{LQ}(0)}{M_W^2} - 2 \frac{s_W}{c_W}
\frac{\Sigma^{\gamma Z}_{LQ}(0)}
{M_Z^2} - \chi_e - \chi_\mu 
\; ,\\
\Delta \kappa^{LQ}_{{univ}}  &=&  - \frac{c_W}{s_W}~ 
\frac{\Sigma^{\gamma Z}_{LQ}(M_Z^2)}{M_Z^2}
- \frac{c_W}{s_W}~ 
\frac{\Sigma^{\gamma Z}_{LQ}(0)}{M_Z^2}
+\frac{c_W^2}{s_W^2} \left[ \frac{\Sigma_{LQ}^Z(M_Z^2)}{M_Z^2}-
\frac{\Sigma_{LQ}^W(M_W^2)}{M_W^2}\right]
\; ,
\end{eqnarray}
where the factors $\chi_\ell$ are defined below.  The leptoquark
contributions to the self-energies can be easily evaluated, yielding
\begin{equation}
{\Sigma}^{V}_{{LQ}}(k^2)  =  - \frac{\alpha_{{em}}}{4\pi} N_c 
\sum_{j} {\cal F}^V_j~
{\cal H} \left ( k^2, M^2\right ) 
\; , \label{sig:g}
\end{equation}
where $N_c = 3$ is the number of colors and the sum is over all
members of the leptoquark multiplet.  The coefficient ${\cal F}^V_j$ is
given by $(Q^\gamma_{j})^2$, $\left ( Q_Z^{j} \right) ^2$, $
-Q^\gamma_{j} Q_Z^{j}$, and $ \left ( T_3^{j} \right )^2/s_W^2$ for $V
= \gamma$, $Z$, $\gamma Z$, and $ W$ respectively.  The function
${\cal H}$ is defined according to:
\begin{equation}
{\cal H}(k^2, M^2) =  - \frac{k^2}{3} \Delta_M - \frac{2}{9}k^2
-  \frac{4 M^2 - k^2}{3} \int^1_0 dx~ \ln \left [
\frac{{ x^2 k^2 - x k^2 + M^2 - i \epsilon}}
{M^2} \right ] \; ,
\end{equation}
with
\begin{equation}
\Delta_M = \frac{2}{4-d} - \gamma_E + \ln(4\pi)  - \ln \left (
\frac{M^2}{\mu^2} \right ) \; ,
\label{delta}
\end{equation}
and $d$ being the number of dimensions.

The factors $\chi_\ell$ ($\ell = e$, $\mu$) stem from corrections to
the effective coupling between the $W$ and fermions at low energy.
Leptoquarks modify this coupling, inducing a contribution that we
parametrize as
\begin{equation}
i \frac{e}{\sqrt{2} s_W}~ \chi_\ell~ \gamma_\mu P_L \; ,
\end{equation}
where $P_L$ ($P_R$) is the left-handed (right-handed) projector and
$\ell$ stands for the lepton flavor.  Since this correction modifies
the muon decay, it contributes to $\Delta r$, and consequently, to
$\Delta \rho_{{univ}}$. Leptoquarks with right-handed couplings,
as well as the $F=0$ ones, do not contribute to $\chi_\ell$. The
analytical for $\chi_\ell$ due to left-handed leptoquarks in the $F=2$
sector can be found in Ref.\ [14].

Corrections to the vertex $Z f \bar{f}$ give rise to non-universal
contributions to $\rho_f$ and $\kappa_f$. We parametrize the effect of
leptoquarks to these couplings by
\begin{equation}
i \frac{e}{2 s_W c_W} \left [ \gamma_\mu F_{VLQ}^{Zf} - \gamma_\mu \gamma_5
F_{ALQ}^{Zf} + I_3^f \gamma_\mu (1 - \gamma_5) \frac{c_W}{s_W} ~
\frac{\Sigma^{\gamma Z}_{LQ}(0)}{M_Z^2} \right ] \; ,
\end{equation}
where for leptons ($\ell$) and leptoquarks with $F=2$
\begin{equation}
\begin{array}{ll}
F^\ell_{VLQ}=  &  \pm F^\ell_{ALQ}=  \frac{g_{LQ,X}^2}{32 \pi^2} N_c
{\displaystyle \sum_{j, q} }
{M^{j}_{\ell q}}^\dagger M^{j}_{q\ell} \\
 &  \left\{ \frac{g^q_X}{2} 
- s_W c_W Q_Z^{j}- \left (g_X^q + 2 s_W c_W Q_Z^{j} \right )~ 
\frac{M^2 - m_q^2}{M_Z^2}
\left [ - \frac{1}{2} \ln \left ( \frac{M^2}{m_q^2} \right ) 
+ \bar{B_0} ( 0, m_q^2,M^2 ) \right ] \right. \\ 
 &  + 2 s_W c_W Q_Z^{j} \frac{M^2 - m_q^2 - \frac{1}{2} M_Z^2}{M_Z^2}
\left [ - \ln \left ( \frac{M^2}{m_q^2} \right ) + \bar{B_0} 
( M_Z^2, M^2, M^2) \right ]  \\
 & + g_X^q \frac{M^2-m_q^2 - \frac{1}{2} M_Z^2}{M_Z^2} \bar{B_0} 
(M_Z^2, m_q^2, m_q^2 ) + g^{\ell}_X \bar{B_1} (0, m_q^2, M^2) \\
  &  + \left [ g_{-X}^q m_q^2 + g_X^q \frac{(M^2-m_q^2)^2}{M_Z^2}
\right ] C_0 (0, M_Z^2, 0, M^2, m_q^2, m_q^2 ) \\
 & \left. - 2 s_W c_W Q_Z^{j} \frac{(M^2-m_q^2)^2 + m_q^2 M_Z^2}{M_Z^2}
C_0 (0, M_Z^2, 0, m_q^2, M^2, M^2) \right\} \; ,
\end{array}
\label{z:ll}
\end{equation}
where the $+$ $(-)$ corresponds to left- (right-) handed leptoquarks
and $g_{L/R}^f = v^f \mp a^f$ with the neutral current couplings being
$a_f = I_3^f$ and $v_f = I_3^f - 2 Q^f s_W^2$.  $M^{j}_{q \ell}$
summarizes the couplings between leptoquarks and fermions.  The
functions $B_1$, $C_0$, $C_{00}$, and $C_{12}$ are the
Passarino-Veltman functions \cite{passa}. We used the convention
$X=L,R$ and $-L=R$ ($-R=L$). We also defined
\begin{eqnarray}
B_0 (k^2, M^2, {M^\prime}^2)  & \equiv &  \frac{1}{2}\Delta_M+
\frac{1}{2} \Delta_{M'} + \bar{B_0} 
(k^2, M^2, {M^\prime}^2 )
\; , \\
B_1 (k^2, M^2, {M^\prime}^2)  & \equiv &  - \frac{1}{2} \Delta_M + \bar{B_1} 
(k^2, M^2, {M^\prime}^2)
\; ,
\end{eqnarray}
with $\Delta_M$ given by Eq.\ (\ref{delta}). From this last expression
we can obtain the effect of $F=2$ leptoquarks on the vertex $Z q
\bar{q}$ simply by the change $\ell \Leftrightarrow q$.  Moreover, we
can also employ the expression (\ref{z:ll}) to $F=0$ leptoquarks
provided we substitute $g_{LQ,X} \Rightarrow h_{LQ,X}$ and $g^q_{\pm
  X} \Rightarrow - g^q_{\mp X}$.

With all this we have
\begin{eqnarray}
\Delta \rho^{LQ}_{{non}}  & = &  \frac{F_{ALQ}^{Zf}}{a_f}(M_Z^2)
\; , \\
\Delta \kappa^{LQ}_{{non}}  & = &  - \frac{1}{2 s_W^2  Q^f} \left [
F_{VLQ}^{Zf}(M_Z^2) - \frac{v_f}{a_f}~ F_{ALQ}^{Zf}(M_Z^2)
\right ]
\; .
\end{eqnarray}

One very interesting property of the general leptoquark interactions
that we are analyzing is that all the physical observables are
rendered finite by using the same counter-terms as appear in the SM
calculations \cite{hollik}. For instance, starting from the
unrenormalized self-energies (\ref{sig:g}) and the mass and
wave-function counter-terms we obtain finite expression for the
two-point functions of vector bosons. Moreover, the contributions to
the vertex functions $Z f \bar{f}$ and $W f \bar{f^\prime}$ are
finite.

In order to check the consistency of our calculations, we analyzed the
effect of leptoquarks to the $\gamma f \bar{f}$ vertex at zero
momentum. It turns out that the leptoquark contribution to this vertex
function not only is finite but also vanishes at $k^2=0$ for all
fermion species.  Therefore, our expressions for the different
leptoquark contributions satisfy the appropriate QED Ward identities,
and leave the fermion electric charges unchanged.  Moreover, we also
verified explicitly that the leptoquarks decouple in the limit of
large $M$.

%**************************************************************

\section{Results and Discussion}
\label{res}

In our analyses, we assumed that the leptoquarks couple to leptons and
quarks of the same family.  In order to gain some insight on which
corrections are the most relevant, let us begin our analyses by
studying just the oblique corrections \cite{obli}, which we
parametrized in terms of the variables $\epsilon_1$, $\epsilon_2$, and
$\epsilon_3$. These variables depend only upon the interaction of
leptoquarks with the gauge bosons and it is easy to see that
leptoquarks contribute only to $\epsilon_2$.  Imposing that this
contribution must be within the limits allowed by the LEP data, 
we find out that the constraints coming from
oblique corrections are less restrictive than the available
experimental limits \cite{lep,ppbar,hera}.

We then performed a global fit to all LEP data including both
universal and non-universal contributions.  In Table \ref{LEPdata} we
show the the combined results of the four LEP experiments \cite{sm}
that were used in our analysis.  In order to perform the global fit we
constructed the $\chi^2$ function associated to these data and we
minimized it using the package MINUIT. We expressed the theoretical
predictions to these observables in terms of $\kappa^f$, $\rho^f$, and
$\Delta r$, with the SM contributions being obtained from the program
ZFITTER \cite{zfit}.  In our fit we used five parameters, three from
the SM: $m_{{top}}$, $M_H$, and $\alpha_s(M_Z^2)$, and two new
ones: $M$, and the leptoquark coupling denoted by $g_{LQ}$.
Furthermore, we have also studied the dependence upon the SM inputs
$M_Z$, $\alpha_{{em}}$, and $G_F$.

%***********
\begin{table}
\caption{LEP data}
\label{LEPdata}
\begin{displaymath}
\begin{array}{|l|l|}
\hline
\hline
\mbox{Quantity}  &  \mbox{Experimental value} \\ \hline
M_Z \mbox{[GeV]}  &  91.1888 \pm 0.0044 \\
\Gamma_Z \mbox{[GeV]}  &  2.4974 \pm 0.0038 \\
\sigma_{\rm had}^0  \mbox{[nb]}  &  41.49 \pm 0.12\\
R_e = \frac{\Gamma({\rm had})}{\Gamma(e^+ e^-)}  &  20.850 \pm 0.067 \\
R_\mu = \frac{\Gamma({\rm had})}{\Gamma(\mu^+ \mu^-)}  &  20.824 \pm 0.059 \\
R_\tau = \frac{\Gamma({\rm had})}
{\Gamma(\mu^+ \mu^-)}  &  20.749 \pm 0.070 \\
A_{FB}^{0e}  &  0.0156 \pm 0.0034 \\
A_{FB}^{0\mu}  &  0.041 \pm 0.0021 \\
A_{FB}^{0\tau}  &  0.0228 \pm 0.0026 \\
A_{\tau}^0  &  0.143 \pm 0.010 \\
A_e^0   &  0.135 \pm 0.011 \\
R_b = \frac{\Gamma(b \bar{b})}{ \Gamma({\rm had})}  & 0.2202 \pm 0.0020\\
R_c = \frac{\Gamma(c\bar{c}) }{\Gamma({\rm had})}  &  0.1583 \pm 0.0098\\
A_{FB}^{0b}  &  0.0967 \pm 0.0038  \\
A_{FB}^{0c}  &  0.0760 \pm 0.0091  \\
\hline
\hline
\end{array}
\end{displaymath}
\end{table}
%%%%%

%***********

\begin{figure}
\epsfxsize=13cm
\begin{center}
%\leavevmode \epsfbox{contournew.ps}
\leavevmode \epsfbox{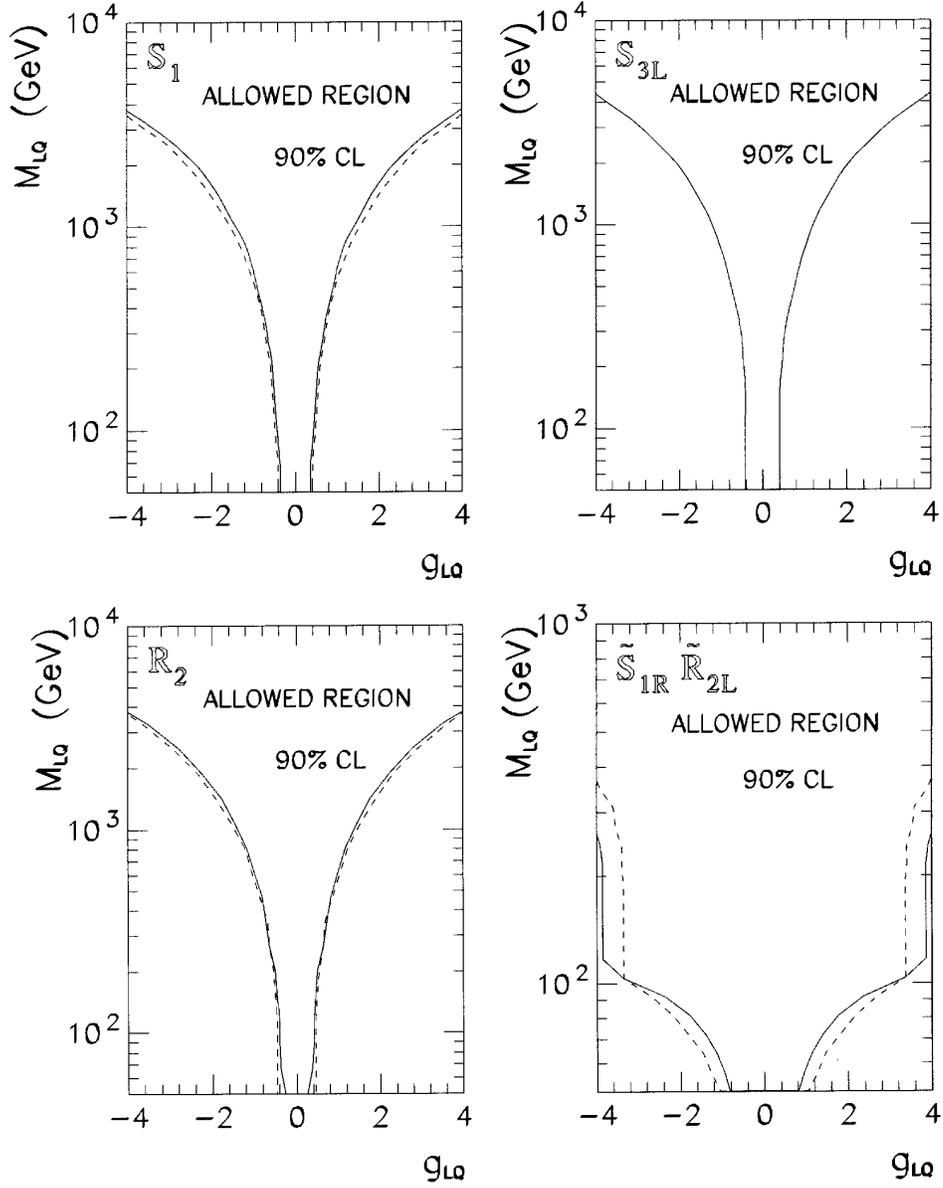}
\end{center}
\caption{Allowed regions (90\% CL) in the plane $M_{LQ}$--$g_{LQ}$ for
third generation leptoquarks. The values of all the other parameters
($m_{{top}}$, $M_H$, $\alpha_s$, $M_Z$, $\alpha_{{em}}$, and $G_F$)
were allowed to vary. The solid lines stand for left-handed leptoquarks
while the dashed ones are for right-handed leptoquarks. Notice the change
of scale in the last window.} 
\label{contours}
\end{figure}

The first part of our analysis consisted of the study of the constraints
on the leptoquark masses and couplings.  In order to determine the
allowed region in the $M_{LQ}$--$ g_{LQ}$ plane, shown in Fig.\
\ref{contours} for the different models, we obtained the minimum
$\chi^2_{{min}}$ of the $\chi^2$ function with respect to the
parameters above for each leptoquark model, and we then  required that $\chi^2
\leq \chi^2_{{min}} +\Delta \chi^2(2,90\% \hbox{CL})$, with
$\Delta\chi^2(2,90\% \hbox{CL})=4.61$.  In this procedure, the
parameters $m_{{top}}$, $M_H$, and $\alpha_s$, as well as the SM
inputs $M_Z$, $\alpha_{{em}}$, and $G_F$ were varied so as to
minimize $\chi^2$.  We must comment here that the dependence on
$\alpha_{{em}}$ and $G_F$ is negligible when they are allowed to
vary in their $90\%$ CL range.  On the other hand, the variation of
$M_Z$ in the interval $91.18\leq M_Z\le 91.196$ leads to a change on
the allowed values of leptoquarks parameters of at most 1\%.

The contour plots exhibited in Fig.\ \ref{contours} were obtained for
third generation leptoquarks. From this figure we can see that the
bounds are much more stringent for the leptoquarks that couple to the
top quark, {\em i.e.} for $S_{1L(R)}$, $S_3$, and $R_{2L(R)}$, since
their contributions are enhanced by powers of the top quark mass.
Moreover, the limits are slightly better for left-handed leptoquarks
than for right-handed ones, given a leptoquark type, and the curve is
symmetric around $g_{LQ}=0$ since the leptoquark contributions are
quadratic functions of $g_{LQ}$.

The contributions from $\tilde R_2$ and $\tilde S_1$ are not enhanced
by powers of the top quark mass since these leptoquarks do not couple
directly to up-type quarks. Therefore, their limits are much weaker,
depending on $m_{{top}}$ only through the SM contribution, and
the bounds for these leptoquarks are worse than the present discovery
limits unless they are strongly coupled ($g_{LQ}^2 = 4 \pi$).
Moreover, the limits on first and second generation leptoquarks are
also uninteresting for the same reason.  Nevertheless, if we allow
leptoquarks to mix the third generation of quarks with leptons of
another generation the bounds obtained are basically the same as the
ones discussed above\footnote{In the case of first generation leptons,
  we must also add a tree level $t$-channel leptoquark exchange to
  some observables.}, since the main contribution to the constraints
comes from the $Z$ widths.

We next present our results as 95\% CL lower limits in the leptoquark
mass and study the dependence of these limits upon all other
parameters. For this, we minimized the $\chi^2$ function for fixed
values of $\alpha_s$, $M_H$, and $m_{{top}}$ and then required
$\chi^2 (\alpha_s, M_H, m_{{top}})\le \chi^2_{{min}}
(\alpha_s, M_H, m_{{top}})+ \Delta\chi^2 (1,90\% \hbox{CL})$,
with $ \Delta\chi^2(1,90\% \hbox{CL})=2.71$. Our results are shown in
Table \ref{res:top} where we give the 95\% CL limits obtained for a
third generation leptoquark for several values of the coupling
constants $g_{LQ}$ ($=\sqrt{4\pi}$, $1$, and $e/s_W$).  The values
given correspond to $m_{top}=175$ GeV and variation of $M_H=60-1000$
GeV and $\alpha_s(M_Z^2)=0.126\pm 0.005$, which is the range
associated to the best values obtained from a fit in the framework of
the SM \cite{sm}.  For a fixed value of $m_{{top}}$ and
leptoquark coupling constant, the dependence on $\alpha_s(M_Z^2)$ and
$M_H$ is such that the limits are more stringent as $\alpha_s(M_Z^2)$
increases and $M_H$ decreases.  The SM parameters $M_Z$,
$\alpha_{{em}}$, and $G_F$ have been also varied in their allowed
range. However, this did not affect the results in a noticeable way.

%*********
\begin{table}
\caption{ Lower limits (95\% CL) for the mass of third generation 
leptoquarks in
GeV for different values of the couplings, assuming $m_{{top}} = 175$
GeV, $\alpha_s(M_Z^2) =  0.126\pm 0.005$, and $M_H = 60-1000$ GeV.}
\label{res:top}
\begin{tabular}{|c|c|c|c|c|c|c|c|}
\hline
\hline
$g_{LQ}$  &  $S_1^R$      &    $S_1^L$     &    $S_3$       &    $R_2^R$     
&    $R_2^L$ &  $\tilde{S}_1^R$  &  $\tilde{R}_2^L$ \\
\hline %\tableline
$\protect\sqrt{4\pi}$  &  5800--3200  &  6000--3500  &  8000--3700 
 &  6000--3300  &  6800--3400  &  300--100  &  
550--120\\
1  &  1200--550  &  1200--600  &  1700--700  &  1250--600  &  1400--600  
&  ---  &  ---\\
${\displaystyle \frac{e}{s_W}}$  &  550--200 &  600--225  &  900--325  & 
 600-250  &  700-250  &  ---  &  --- \\
\hline
\hline
\end{tabular}
\end{table}

%*********

We would like to stress that the large apparent uncertainty associated
with the value of $\alpha_s$ and $M_H$ can be considered somehow
fictitious as the value of $\chi^2_{{min}}$ grows very fast when we
move from the central value $\alpha_s=0.126$, $M_H=300$ GeV what means
that the quality of the fit for the extreme values of these parameters
is rather bad. For instance, $\alpha_s=0.117$, results in a too high
$\chi^2$, even in the context of the SM ($\chi^2_{{min}}>26/12$).

%******************************************************************************

\section{Acknowledgements}

I would like to thank Alan Sommerer for his hospitality.  This work
was supported by Conselho Nacional de Desenvolvimento Cient\'{\i}fico
e Tecnol\'ogico (CNPq) and by Funda\c{c}\~ao de Amparo \`a Pesquisa do
Estado de S\~ao Paulo (FAPESP).

%******************************************************************************

%******************************************************************************
%Figures
%%%%%%%%%

%**********
% TABLES
%**********

%%%%%

\protect

%******************************************************************************

\end{document}